\documentclass[preprint2]{aastex}

\shorttitle{Three Mistakes in Pulsar Electrodynamics}
\shortauthors{Z. X. Liang and Y. Liang}

\begin{document}

\title{Three Mistakes in Pulsar Electrodynamics}

\author{Zhu-Xing Liang,  Yi Liang}
\affil{KPT lib, Shuixiehuadu18-4-102, Zhufengdajie, Shijiazhuang,
Hebei, China 050035} \email{zx.liang55@gmail.com}



\begin{abstract}
 In the paper \emph{Pulsar Electrodynamics}, published in 1969,
  Goldreich and Julian propose some basic properties
  of pulsars, such as the oft-cited Goldreich-Julian density,
  light cylinders, open and closed magnetic field lines,
  corotation and so on. However, inspection of their
  mathematics reveals three mistakes: first, the relative velocity
   is irrelevantly replaced by the corotation velocity; second,
   a hypothesis in their theory  is contradictory to Maxwell's equations;
   and third, their theory neglected a special solution of
   the frozen-in field equation which is of particular importance
   for pulsar research. We additionally describe the results of
   a series of magnetohydrodynamic experiments that may be beneficial
   to the understanding of pulsar electrodynamics.

\end{abstract}

\keywords{(stars:) pulsars: general, (magnetohydrodynamics) MHD,
stars: magnetic fields}
\section{Introduction}
Soon after pulsars were discovered, \citet{GJ69} (hereafter GJ69)
discussed their electrodynamics, and proposed a series of concepts
that have remained in use ever since \citep{LG06}. These include the
Goldreich-Julian density, light cylinders, open and closed magnetic
field lines, corotation and so on. However, inspection of their
mathematics reveals three mistakes, which are discussed in Section 2
of the present paper. In Section 3, we describe the results of a
series of magnetohydrodynamic experiments that may be beneficial to
the understanding of pulsar electrodynamics.

GJ69 discussed only the problem of aligned pulsars. Thus, our
discussion here will also be focused on the aligned cases.
\section{Three Mistakes in the theory}
\subsection{Relative Velocity and Corotation Velocity}
In GJ69, the basis of all calculations is the equation
\begin{equation}
\emph{\textbf{E}}+\frac{\emph{\textbf{v}}}{c}\times\emph{\textbf{B}}=0.
\label{}
\end{equation}
The second term of equation (1) is the induced electric field
(corresponding to the Lorentz force), and it is widely recognized
that \textbf{\emph{v}} is the relative velocity of a charged
particle traveling through the magnetic field. If the plasma and
charged particles of the pulsar cannot travel perpendicularly to the
magnetic field because of the effects of frozen-in field lines, we
find that
$\frac{\emph{\textbf{v}}}{c}\times\emph{\textbf{B}}\equiv0$ and
$\emph{\textbf{E}}\equiv0$, sequentially, so that Equation (1) is
meaningless.

The first equation in GJ69 is
\begin{equation}
\emph{\textbf{E}}+\frac{\Omega\times
\emph{\textbf{r}}}{c}\times\emph{\textbf{B}}=0, \label{}
\end{equation}
where $\Omega$ is the angular velocity of the pulsar. Obviously,
replacing relative velocity $\emph{\textbf{v}}$ by $\emph{\textbf{$
\Omega\times$r}}$   means that the relative velocity of a charged
particle traveling through the magnetic field is equal to
$\emph{\textbf{$ \Omega\times$r}}$, that is,
$\emph{\textbf{v}}=\emph{\textbf{$ \Omega\times$r}}$. In this case
the particle corotates with the pulsar and the magnetic field is
stationary. However, GJ69 state that the charged particles are
threaded by magnetic field lines and corotate with the star,
implying that the charged particles cannot travel perpendicularly to
the magnetic field lines.

Therefore, the authors' statements are self-contradicting.

\subsection{Rotation of the Magnetic Field}

GJ69 propose the hypothesis that when a pulsar rotates, the magnetic
field lines corotate rigidly with the star's angular velocity.
Whether or not the magnetic field lines corotate with its magnetic
dipole  while the latter is rotating around its own magnetic axis, a
problem known as unipolar induction, has been debated for a century.
Some researchers believe that they rotate together (the M
hypothesis), while others think they do not (the N hypothesis).
Here, we refute the M hypothesis by showing that it contradicts
Maxwell's equations.

\begin{figure}[ht]
\epsscale{1} \plotone{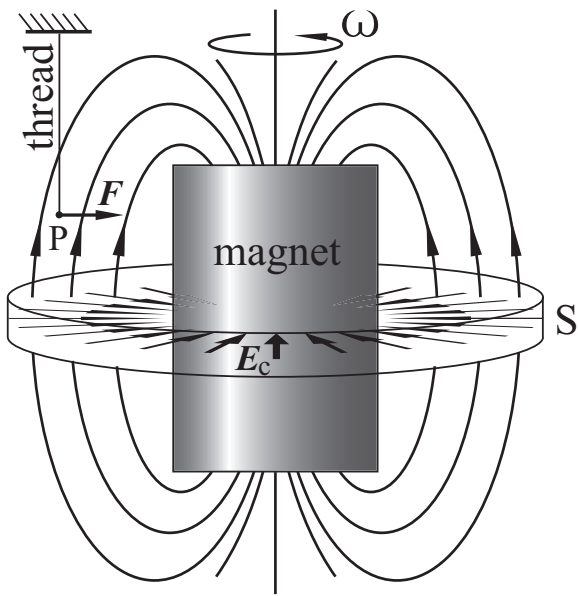} \caption{Electric force on the
testing charge. When the magnet rotates, does the electric force
$\textbf{\emph{F}}$ appear? The M hypothesis and the Maxwell's
equation will give the opposite answers to this question.
\label{Fig1}}
\end{figure}

In the light of the inference of the M hypothesis, when the magnet
in Fig.~\ref{Fig1} rotates, the magnetic field lines rotating with
the magnet together will apply an electric force $\emph{\textbf{F}}$
on the testing charge P suspended in the magnetic field. However, as
we know, the electric field $\emph{\textbf{E}}$ anywhere can be
described as
\begin{equation}
\emph{\textbf{E}} = \emph{\textbf{E}}_c + \emph{\textbf{E}}_i,
\label{}
\end{equation}
where $\emph{\textbf{E}}_c$ is the electric field energized by the
charges and $\emph{\textbf{E}}_i$ is the vortex electric field
induced by the change of magnetic field. The following equations can
be derived from Maxwell's equations:
\begin{equation}
\left\{
\begin{array}{l}
\nabla \cdot  \textbf{\emph{E}}_c  = \rho/\varepsilon _0\;
\left(\textup{or} \;  \oint_{\,S}\textbf{\emph{E}}_c \cdot
\mathrm{d} \textbf{\emph{S}} = Q/\varepsilon _0\right),\\ \nabla
\times \textbf{\emph{E}}_c  = 0,
\end{array}\right.
\label{}
\end{equation}
and
\begin{equation}
\left\{
\begin{array}{l}
\nabla  \times \emph{\textbf{E}}_i  =  - \,\partial
\emph{\textbf{B}}/\partial t,\\ \nabla  \cdot \textbf{\emph{E}}_i
= 0.
\end{array}\right.
\label{}
\end{equation}
If there is an electric field $\emph{\textbf{E}}_c$ when a magnet
rotates, it is certain that this electric field will be
symmetrically distributed around the axis of the magnet (shown in
Fig 1). As long as the quantity of electric charge in the closed
surface S is zero, that is, Q=0, Equation (4) implies the electric
field $\emph{\textbf{E}}_c$ is zero. Therefore, provided the magnet
has no charges, no matter whether it rotates or not, the electric
field $\emph{\textbf{E}}_c$ will always be zero. On the other hand,
whether the magnet rotates or not, $\emph{\textbf{B}}$ is always a
constant everywhere, while according to Equation (5), the curl of
the vortex electric field $\emph{\textbf{E}}_i$ is zero and clearly
$\emph{\textbf{E}}_i$ also vanishes. In other words, according to
Maxwell's equations, no matter whether the magnet rotates or not,
there is no electric field $\emph{\textbf{E}}$ in the outside space
around the magnet and it is impossible to apply an electric field
force $\emph{\textbf{F}}$ on the charge P. However, the M hypothesis
infers that the charge P can be acted upon by the electric field
force, thus contradicting Maxwell's equations. Therefore, we suggest
the M hypothesis is flawed and that the N hypothesis should be
applied to the study of astrophysics.

Unfortunately, the M hypothesis has had an important influence on
the pulsar theory. Current consensus in the scientific community
appears to be that general physicists prefer the N hypothesis,
whereas almost all astrophysicists prefer the M hypothesis. We
therefore believe it is worthwhile to discuss the nature of the
magnetic field in more detail, to highlight the shortcomings of the
M hypothesis. For this reason, we designed a set of cartoons
depicting the \emph{moving characteristics} of the magnetic field,
which are available for viewing at
http://www.pulsar2.com/English/fieldfilm1.htm.

The corotation model proposed in GJ69 is based on the M
hypothesis, which contradicts Maxwell's equation. Therefore, it
cannot be valid.

\subsection{Effect of Frozen-in Field Lines}

The point of view of GJ69, based on the so-called effect of
frozen-in field lines deduced from magnetohydrodynamics, is that
the magnetosphere particles cannot travel perpendicularly to the
field lines. Here we will show that this concept is untenable.

\begin{figure}[ht]
\epsscale{1} \plotone{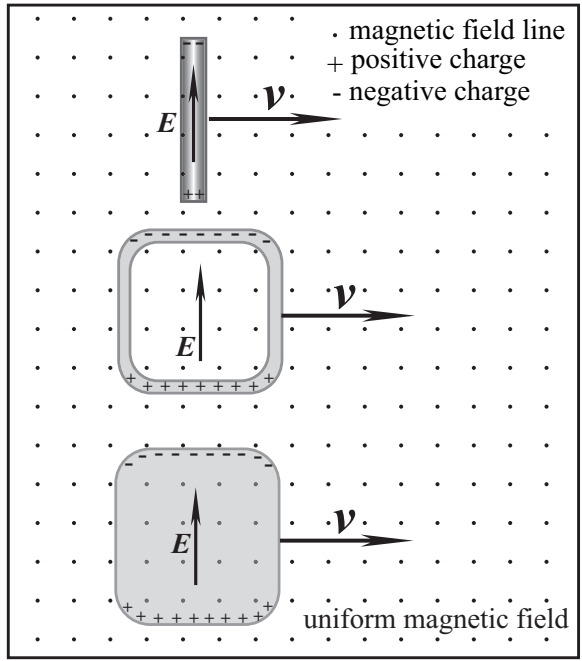} \caption{The frozen-in effect.
Whether the ideal conductive fluid can be frozen with the magnetic
field lines together or not when moving in the uniform magnetic
field, the magnetohydrodynamics holds the opposite opinion from the
general physics. The mistake of the magnetohydrodynamics is that it
irrelevantly omitted the existence of the polarized charges and
electric field.\label{Fig2}}
\end{figure}
\begin{figure}[ht]
\epsscale{1} \plotone{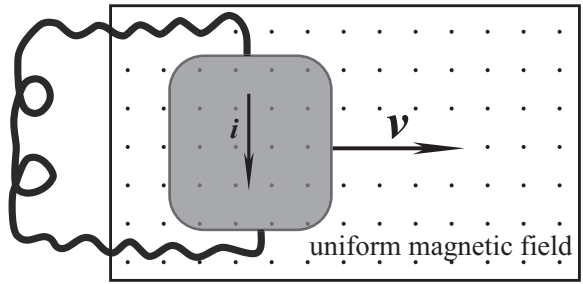} \caption{The influence of boundary
condition on the frozen-in effect. Only when the polarized charges
of the boundary can continuously flow away and the current
\textbf{\emph{i}} forms, the frozen-in effect will occur.
\label{Fig3}}
\end{figure}
\begin{figure}[ht]
\epsscale{1} \plotone{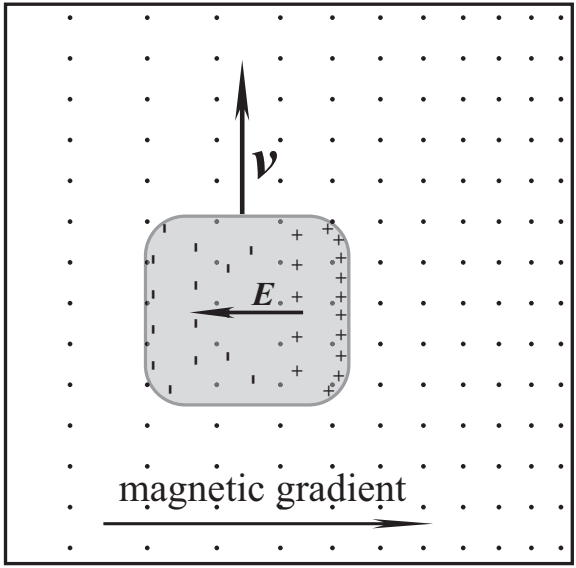} \caption{The second non-frozen
situation. Even if the conductive fluid moves in the nonuniform
magnetic field, as long as it moves along an isomagnetic surface,
the frozen-in effect will not occur. \label{Fig4}}
\end{figure}
\begin{figure}[ht]
\epsscale{1} \plotone{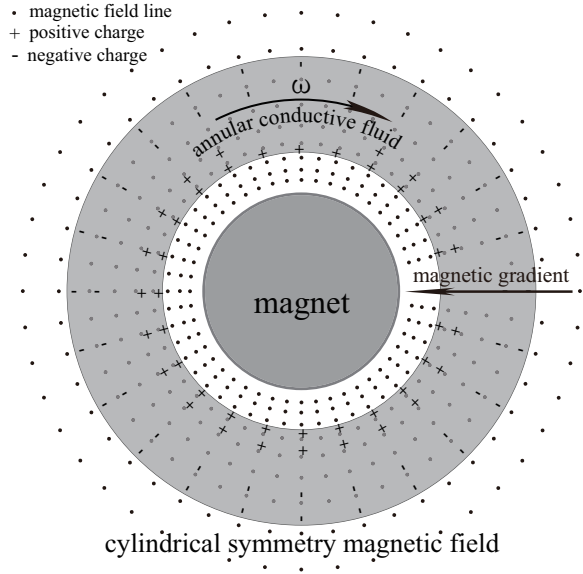} \caption{The third non-frozen
situation. In the cylindrically symmetrical magnetic field, the
rotation of the conductive fluid around the magnetic axle does not
lead to the generation of the magnetic frozen phenomenon either,
because this induced electric field has zero curl, which is just the
same situation in the vicinity of pulsar's equator. \label{Fig5}}
\end{figure}

According to general physics, a wire, annular wire or plate made up
of conducting material (Fig. 2), when moving with constant velocity
in a uniform magnetic field and cutting the magnetic field lines,
can neither produce electric current, nor receive any resistance
from the magnetic field, so that the conductors cannot be frozen
together with the magnetic field. This conclusion is independent of
both the kind and the conductivity of materials, even if a
conductive fluid such as mercury or plasma is used to make the
objects shown in Fig.2, so the frozen field phenomenon cannot exist.
However, the magnetohydrodynamical theory states that an ideal
conductive fluid is frozen in the magnetic field and that it cannot
travel across the field lines. Thus, there is a contradiction.

We have found that in all studies of the frozen-in effect, the
polarization charges and the electric field\emph{ \textbf{E}} shown
in Fig. 2 were irrelevantly disregarded. It is the electric field
\emph{\textbf{E}} that thaws the frozen phenomenon and can be simply
calculated using the drift theory of charged particles, whereby the
drift velocity driven by the electric field\emph{ \textbf{E}} is
equal to the advection velocity of the fluid \emph{\textbf{v}}, that
is to say $\emph{\textbf{v}}\!_{shift} \equiv \emph{\textbf{v}}$.
The direction of the drift velocity $\emph{\textbf{v}}\!_{shift}$ is
always perpendicular to both the electric field and the magnetic
field. Therefore, the conductive fluid can cross a uniform magnetic
field without any resistance and cannot be frozen together with the
magnetic field lines.

If a wire is connected to the two ends of the conductor and the
charges are allowed to flow out, as shown in Fig. 3, the current
\emph{\textbf{i}} and the frozen-in field effect will appear and
the magnetic field will exert a resistance on the conductor.
Consequently, the existence of the frozen phenomenon in a uniform
magnetic field is a function of the boundary conditions of the
fluid. If the charges can continuously flow, with current
\emph{\textbf{i}}, the frozen phenomenon can be manifested.
Otherwise, there is no frozen phenomenon.

An ideal conductive fluid can cross not only a uniform magnetic
field, but also a non-uniform field, still free from any resistance,
providing the fluid moves along an isomagnetic surface of the
magnetic field as shown in Fig. 4. This conclusion can be obtained
from the frozen-in field equation:
\begin{equation}
\frac{\partial\mathbf{\emph{\textbf{B}}}}{\partial{t}}=\nabla\times
(\emph{\textbf{v}}\times\emph{\textbf{B}}) , \label{e4}
\end{equation}
where \emph{\textbf{v}} is the velocity of the conductive fluid
elements moving through the magnetic field, and
\emph{\textbf{v}}$\times$\emph{\textbf{B}} is the induced electric
field.

As we know, the essential condition under which a force can exist
between the conductive fluid and magnetic field is that of the
existence of an eddy current inside the conductive fluid, and the
essential condition for the existence of an eddy current appearing
is an eddy induced electric field. When both sides of Equation (6)
are equal to zero, the induced electric field inside the conductive
fluid is an irrotational field. In this case, there is neither an
eddy electric field nor an eddy current in the conductive fluid.
Sequentially, there is no magnetic force acting on the conductive
fluid when it passes through the magnetic field with velocity
\textbf{\emph{v}}. This case is a special solution of the frozen-in
field equation (6).

The frozen-in field equation (6) can be expanded as:
\begin{equation}
\frac{\partial\mathbf{\emph{\textbf{B}}}}{\partial{t}}=
\textbf{\emph{v}}(\nabla\cdot\emph{\textbf{B}})
-\textbf{\emph{B}}(\nabla\cdot\emph{\textbf{v}})
-(\textbf{\emph{v}}\cdot\nabla)\emph{\textbf{B}}
+(\textbf{\emph{B}}\cdot\nabla)\emph{\textbf{v}}.\label{}
\end{equation}

 The first term on the right-hand side includes the divergence of
the magnetic field $\nabla\cdot\emph{\textbf{B}}$, therefore, this
term is always zero.

The factor $\nabla\cdot\emph{\textbf{v}}$ in the second term is the
divergence of the velocity field. When the fluid moves without
expansion or compression (i.e., the bulk has not change), this term
is also zero.

When the path of each conductive fluid element follows an
isomagnetic surface (i.e., constant B), the third term
$(\textbf{\emph{v}}\cdot\nabla)\emph{\textbf{B}}$ is also equal to
zero.

If all of the fluid elements threaded by a magnetic field line have
the same velocity, without differential movement, the fourth term
$(\textbf{\emph{B}}\cdot\nabla)\emph{\textbf{v}}$ is also zero. If
the shapes of magnetic field lines have not change and this term is
not equal to zero, the shape of the fluid will change.

Therefore, providing the shape and the size of the conductive fluid
remain the same, all of the situations shown in Fig. 2, Fig. 4 and
Fig. 5 correspond to a special solution where all four terms are
zero in the right-hand side of equation (7). In these cases the
induced electric field inside the conductive fluid are irrotational,
and the frozen-in phenomenon is not manifested. Because the case
illustrated in Fig. 5 is equivalent to that in the vicinity of a
pulsar's equator, the viewpoint that the plasma of a pulsar cannot
cross the magnetic field lines is inaccurate.

Another description of the frozen-in field equation is
\begin{equation}
\frac{d\Phi}{dt}=0. \label{e6}
\end{equation}
From this equation, one can see that the essence of the frozen-in
field theorem is the conservation of magnetic flux inside the fluid.
This conservation principle does not imply that the conductive fluid
cannot pass through magnetic field lines, only that the magnetic
field lines entering a given fluid element are equal in number to
those leaving the element at all times.

\begin{figure}[ht]
\epsscale{1} \plotone{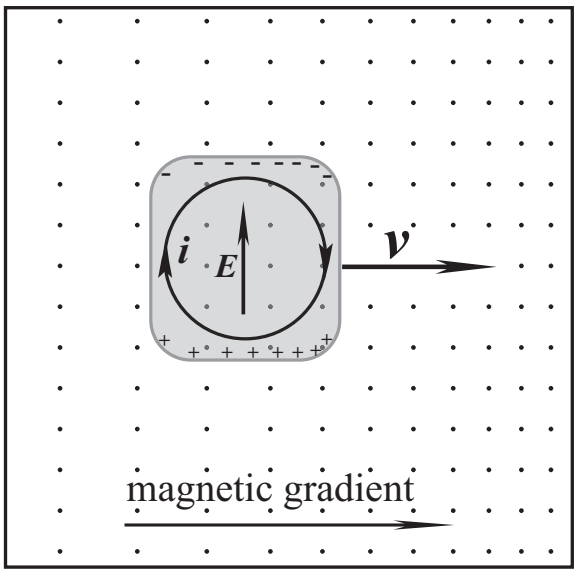} \caption{A frozen-in situation. The
movement of the conductive fluid, if the velocity has a component in
the direction of the magnetic gradient, will lead to the emergence
of the frozen-in effect. \label{Fig6}}
\end{figure}

\begin{figure}[ht]
\epsscale{1} \plotone{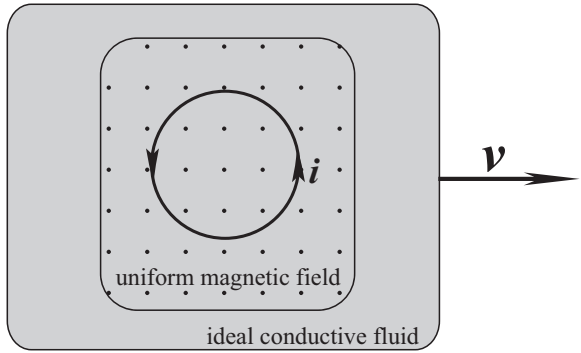} \caption{Another frozen-in
situation. If the size of the magnetic field is smaller than the
size of the conductive fluid, even if the magnetic field is uniform,
the frozen-in effect will also occur, because the magnetic field has
a high gradient near its boundary. \label{Fig7}}
\end{figure}

On the condition that the shape and the size of the conductive fluid
do not change, two frozen examples are respectively shown in Fig. 6
and Fig. 7. In general, if the magnetic field intensity in the
direction of travel of the conductive fluid changes, the conductive
fluid will be acted upon by the frozen resistance.

Further, when an individual charged particle crosses a magnetic
field the Lorentz force will compel it to gyrate around the magnetic
field lines. However, within the conductive fluid, the appearances
of polarization charges and a polarization electric field create a
very different situation. Under some special conditions such as
those illustrated in Figs. 2, 4 and 5, the polarized electric field
can counteract the Lorentz force and allow the conductive fluid to
cross the magnetic field without any resistance.

\section{Magnetohydrodynamic experiments}
To illustrate the above analysis, we conducted a series of
magnetohydrodynamic experiments.

 The main materials used in our experiments were liquid mercury
 and solid magnets. The mercury simulated the plasma, while the magnet
 simulated the pulsar. For the
sake of simplicity and clarity, our experiments were recorded as
videos which are available at YouTube. Briefly, the mercury is
placed in a cylindrical trough surrounding a rotating platform
with a receptacle for the solid magnet.

The experiments were divided into two subgroups. The first consisted
of three pairs of experiments whose purpose was to ascertain the
conditions under which the magnet can drive rotation in the
surrounding mercury. In each experiment our basis for comparison was
an \emph{aligned rotator}\,: a cylindrically symmetric magnet whose
magnetic axis is the same as the axis of rotation. The three
experiment pairs were as follows:
\begin{enumerate}
      \item  Drive Experiment A shows the contrast between
      the aligned rotator and a rotator whose magnetic axis
      is orthogonal to the axis of rotation. The URL is

      http://www.youtube.com/watch?v=YCc4ybfKiY0.

      \item  Drive Experiment B shows the contrast between
      the aligned rotator and a magnet whose magnetic axis
      is parallel to but displaced from the axis of rotation.
      This video can be found at

      http://www.youtube.com/watch?v=ZiVNxVAqUKc.

      \item Drive Experiment C shows the contrast between the aligned rotator
      and a magnet which is not cylindrically symmetric.
      The video is at URL

      http://www.youtube.com/watch?v=SBAzUBzc2xM.
\end{enumerate}
The first subgroup of experiments demonstrates that a cylindrically
symmetric magnetic field\footnote{A large number of discussions in
this paper are carried on under the condition of cylindrical
symmetry. However, in the experiments, though the magnetic field
generated by the cylindrical magnet doesn't completely tally with
the condition of cylindrical symmetry, the middle of magnetic field
well similarly tallies with the cylindrical symmetry. Therefore,
most discussions are limited to the middle of magnetic field.}cannot
drive the mercury when its magnetic axis is the same as the axis of
rotation. When the magnetic axis is unaligned with the spin axis or
the distribution of the magnetic field is not cylindrically
symmetric, however, the mercury can be driven.

In the second subgroup of experiments, we ascertain the conditions
under which rotating mercury can be slowed down by a stationary
magnet. To do so, we spun the mercury trough rather than the
central magnet until the mercury's revolution reached a steady
state. In each experiment our basis of comparison was again an
\emph{aligned magnet}\,: a cylindrically symmetric magnet whose
magnetic axis was the same as the axis of rotating mercury. With
the aligned magnet, the mercury eventually revolved at the same
rate as the trough. In the other three cases, braking forces
limited the mercury to a slower rate. We used the same magnets in
every pair of experiments seen in Braking Experiments A, B and C.
\begin{enumerate}
    \item Braking Experiment A shows the contrast between the aligned
    magnet and the orthogonal magnet, and can be viewed at

    http://www.youtube.com/watch?v=C7DxRoVdtxw.

    \item Braking Experiment B shows the contrast between the aligned
    magnet and the off-axis magnet, and can be viewed at

    http://www.youtube.com/watch?v=p9UzfT6UHcE.

    \item Braking Experiment C shows the contrast between the aligned
    magnet and the magnet which is cylindrically asymmetric,
    and can be viewed at

    http://www.youtube.com/watch?v=Eq0B5XR\_o8U.

\end{enumerate}
These experiments demonstrate that a coaxial, cylindrically
symmetric magnetic field has no braking effect on the mercury. In
other words, under some special conditions the mercury can cross the
magnetic field lines without experiencing a braking force. The
magnet can exert a braking effect on the mercury only when its
magnetic axis is unaligned with the axis of mercury rotation or when
the magnetic field is not cylindrically symmetrical.

\section{Discussion}

The experimental results reveal two phenomena worthy of special
attention:
\begin{enumerate}
    \item A rotating magnet cannot drive the stationary mercury when the axes of
    rotation of the mercury and the magnet are identical. This result is apparent based
    on the N hypothesis, because the stationary magnetic field lines are unable to induce rotation in the mercury.
    \item A stationary magnet cannot slow down revolving mercury
    when the axis of rotation is identical to the magnetic axis. This result
    is apparent based on the analytical results in Section 2.3 and Fig. 5.
\end{enumerate}
In summary, no exchange of angular momentum will occur between the
magnet and the mercury under conditions where the magnetic field
and the mercury share cylindrical symmetry. The degrees of axial
alignment and symmetry thus control the exchange of the angular
momentum. Of course, a change of magnetic flux through each fluid
element is a necessary condition for the exchange of angular
momentum.

 The plasma surrounding a pulsar covers the entire spherical surface. In our
experiments, the mercury circulates within a region near the
equatorial plane. Therefore, the results of our study are only
directly comparable to the behavior of plasma near the equatorial
plane.

The above discussion refers only to aligned pulsars. Pulsars with
unaligned magnetic fields are more complicated. As shown in \\
http://www.pulsar2.com/English/fieldfilm7.htm, while the magnetic
axis of an oblique rotator swings around, the magnetic field lines
do not rotate around the magnetic axis. When the magnetic dip angle
is small, the magnetic field lines near the equatorial plane can be
considered approximately stationary. Thus, our results are still
noteworthy.

Results presented by both GJ69 and the present paper are based on
the theory of classical mechanics without considering relativistic
effects in the plasma. Even if it is necessary to consider
 these effects, we can make relativistic corrections to the
classical mechanics, but the laws of classical mechanics can never
be violated. However, ideas proposed by GJ69 contradict the laws
of classical mechanics.

\section{Conclusions}

The following conclusions can be drawn from our experiments and
analysis:

\begin{enumerate}
\item Three mistakes were made by GJ69: first, the relative
velocity is irrelevantly replaced by the corotation velocity;
second, their theory is based on the M hypothesis which is
contradictory to Maxwell's equations; and third, their theory
neglected a special solution of the frozen-in field equation which
is of particular importance for pulsar research.

\item The magnetic field near the equatorial plane of an aligned pulsar is steady,
so it can neither drive nor brake the plasma. Thus, the viewpoint
that magnetosphere particles can only move along magnetic field
lines is wrong and the corotation model is flawed.

\item The corotation model should be abandoned and the models of pulsar electrodynamics
should be reformulated under new assumptions.

\end{enumerate}

\end{document}